\documentclass[a4paper,11pt]{article}
\usepackage{pos}
\usepackage{subfig}

\newcommand{\GeV}{\ensuremath{\text{Ge\kern -0.1em V}}}
\newcommand{\TeV}{\ensuremath{\text{Te\kern -0.1em V}}}
\title{Observation of quantum entanglement in top quark pairs at the ATLAS experiment}

\author*[a]{Baptiste Ravina}

\affiliation[a]{II. Physikalisches Institut, Georg-August-Universit\"{a}t G\"{o}ttingen,\\
  Friedrich-Hund-Platz 1, 37077 G\"{o}ttingen, Germany}

\emailAdd{baptiste.ravina@cern.ch}

\abstract{The ATLAS Collaboration has recently reported the first observation of quantum entanglement in top quark pair production at the Large Hadron Collider (LHC).
A brief review of the experimental status of top quark pair spin correlations is given, before highlighting the expected differences between classical and quantum correlations.
A criterion for the detection of spin entanglement in top quark pairs has been proposed; the experimental strategy to implement it and observe the effect is discussed, and the positive results of the ATLAS measurement are presented.
}

\FullConference{12th Large Hadron Collider Physics Conference  (LHCP2024)\\
3-7 June 2024 \\
Boston, USA \\}

\begin{document}
\maketitle

\section{Introduction: spin correlations}

The top quark is the heaviest known elementary particle of the Standard Model (SM) of particle physics.
It is also singled out by its exceptionally short lifetime $\tau_t\sim 5\times 10^{-25}$s, much smaller than the typical timescale of quantum chromodynamcis (QCD) $1/\Lambda_\text{QCD}\sim 10^{-24}$s.
As such, the top quark evades the processes of hadronisation and spin decorrelation, and provides a unique opportunity to study the properties of a ``bare'' quark~\cite{Bigi:1986jk}.

The CKM matrix element $V_{tb}$ being very close to unity, the top quark can be assumed in the SM to decay exclusively via the $t\to Wb$ process, featuring a maximally parity-violating weak interaction.
The unaltered spin information of the top quark is transferred to the bottom quark and the $W$ boson decay products, which in turn are emitted in preferential directions depending on their helicity.
At the Large Hadron Collider (LHC), top quarks are produced mainly in top-antitop pairs ($t\bar{t}$) through gluon fusion, leading to significant spin correlations of the top and antitop quarks.
These correlations can then be observed in the angular distributions of the decay products, after a suitable choice of basis.

These effects are summarised in the doubly-differential cross section of the $t\bar{t}$ decay process, in terms of the solid angles $\Omega_\pm$ of the decay products of the (anti-)top quarks~\cite{Bernreuther:2015yna}:
\begin{equation}\label{eq:spincorrs}
  \dfrac{1}{\sigma}\dfrac{\mathrm{d}\sigma}{\mathrm{d}\Omega_+\mathrm{d}\Omega_-}=\dfrac{1}{4\pi^2} \left( 1 + \alpha_+\mathbf{B_+}\cdot \hat{q}_+ + \alpha_-\mathbf{B^-}\cdot \hat{q}_- + \alpha_+\alpha_-\hat{q}_+\cdot\mathbb{C}\cdot\hat{q}_- \right),
\end{equation}
where $\mathbf{B_\pm}$ are the polarisation vectors of the (anti-)top quarks, which are zero at leading order (LO) in QCD, and $\mathbb{C}$ is the top spin correlation matrix.
The spin analysing powers $\alpha_\pm$ represent the degree of alignment of direction of flight of the (anti-)top quark decay products $\hat{q}_\pm$ (taken in the rest frame of their parent top) and the actual spin orientation of the (anti-)top quarks.
The spin analysing power is maximal for charged leptons, $\lvert\alpha_\ell\rvert\sim 1$; this motivates the study of spin correlations in the dileptonic $t\bar{t}$ final state.

\begin{figure}[!htb]
  \centering
  \subfloat[\label{fig:atlas_dphill}]{\includegraphics[width=0.42\textwidth]{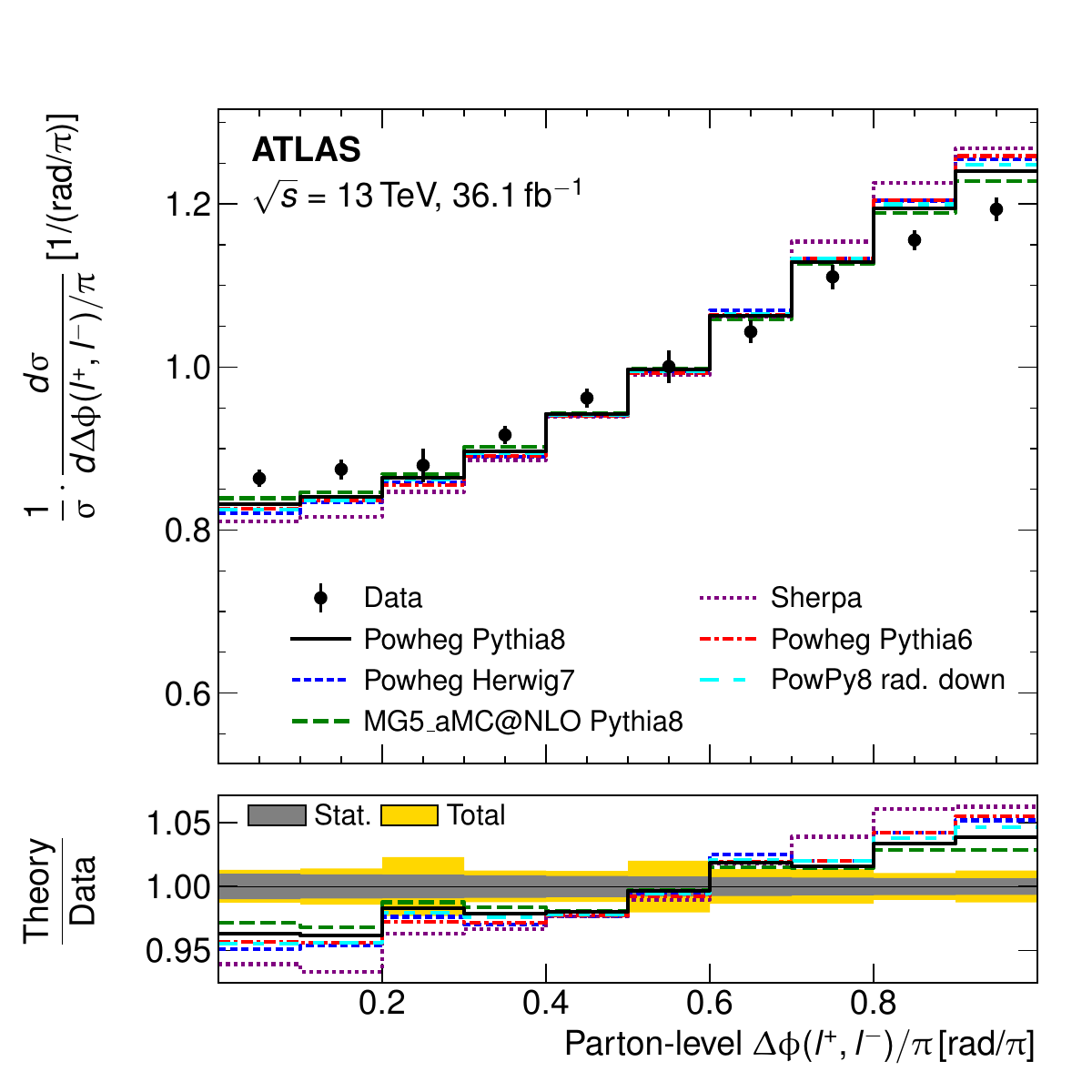}}
  \subfloat[\label{fig:cms_spincoefs}]{\includegraphics[width=0.42\textwidth]{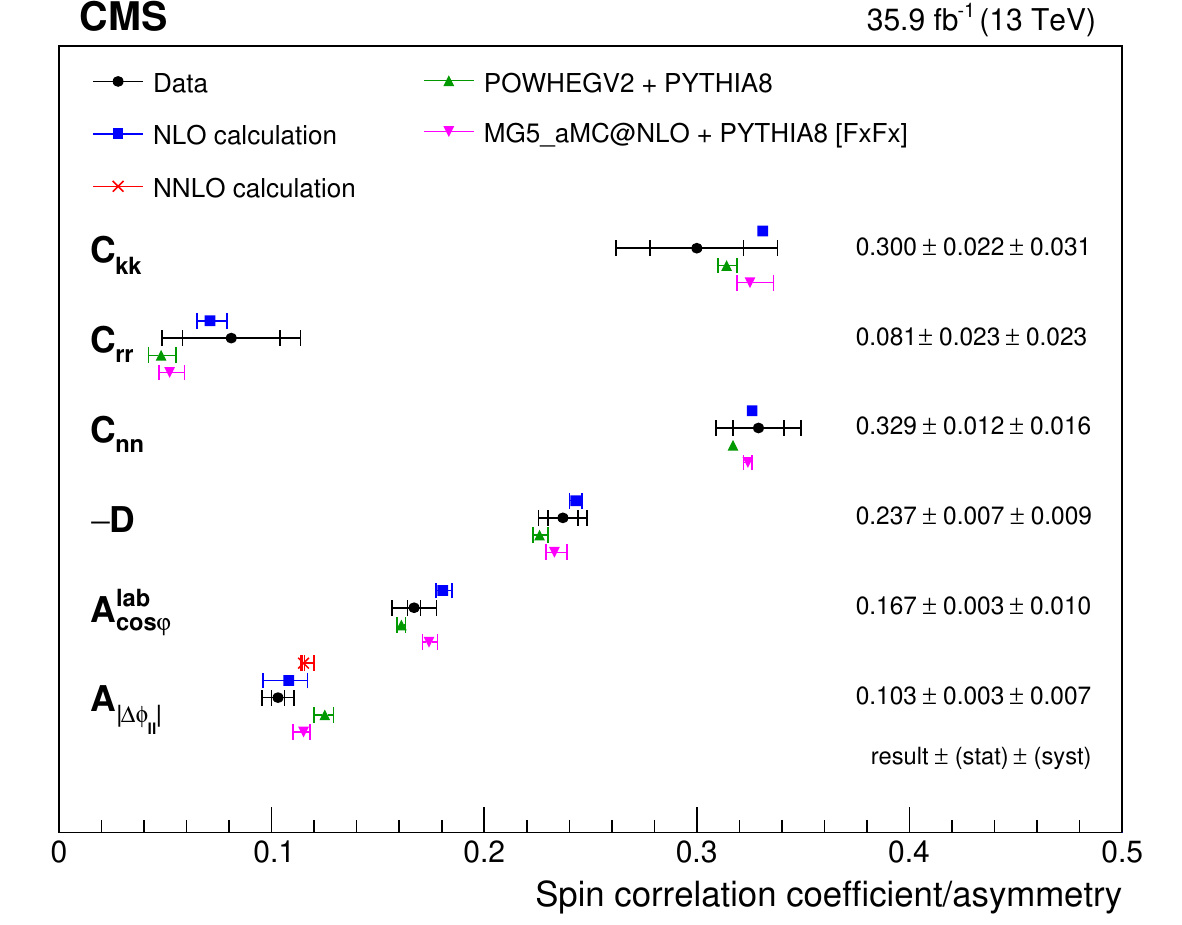}}
  \caption{(a) Normalised parton-level differential cross section of $\Delta\phi(\ell^+,\ell^-)$~\cite{ATLAS:2019zrq}.
  (b) Measured values of the spin coefficients and asymmetries~\cite{CMS:2019nrx}.
  The results are compared to SM predictions at NLO and NLO+PS accuracy.}
\end{figure}

Back in 2020, the experimental state-of-the-art for top spin correlations at the LHC was given by two sets of results: a measurement of spin-sensitive leptonic observables by the ATLAS Collaboration~\cite{ATLAS:2019zrq}, and a direct extraction of the $t\bar{t}$ spin density matrix by the CMS Collaboration~\cite{CMS:2019nrx}.
Figure~\ref{fig:atlas_dphill} shows the azimuthal separation (in the laboratory frame) between the two charged leptons in the dileptonic $t\bar{t}$ final state, unfolded to parton-level by the ATLAS Collaboration: this distribution is characterised by mostly back-to-back leptons (large cross section at high $\Delta\phi(\ell^+, \ell^-)$), but also by an enhancement of collinear lepton production (at low $\Delta\phi(\ell^+, \ell^-)$).
In fact, the data are observed to prefer an even stronger degree of spin correlations than predicted by the SM, and are seemingly in tension with the various Monte Carlo (MC) and parton shower (PS) event simulations.
This tension was later resolved by dedicated next-to-next-to-leading-order (NNLO) QCD calculations~\cite{Behring:2019iiv}.
The CMS results in Figure~\ref{fig:cms_spincoefs} further establish the presence of spin correlations in dileptonic $t\bar{t}$ events: the diagonal elements of the $\mathbb{C}$ matrix are significantly non-zero, and in general display good agreement with the SM predictions.

\section{Classical versus quantum correlations}

The award of the Nobel prize in Physics 2022 to Aspect, Clauser, and Zeilinger, for their experimental demonstration of quantum entanglement with photons, the violation of Bell's inequalities, and pioneering work in quantum information science, has renewed the interest of the public in these fundamentally quantum phenomena.

In the wake of the aforementioned ATLAS and CMS results, the authors of Ref.~\cite{Afik:2020onf} set out to find how spin entanglement would differ phenomenologically from the classical description of spin correlations in $t\bar{t}$ events, and whether this effect could be observed at the LHC.
They proposed the first analysis of top quark production from the point of view of quantum information theory, treating it as a bi-partite qubit system, and following well-established recipes to derive a condition for entanglement.
Starting from the spin density matrix associated to Equation~\eqref{eq:spincorrs} and applying the Peres-Horodecki criterion~\cite{Peres:1996dw,Horodecki:1997vt} leads to the equivalence relations for entanglement
\begin{equation}\label{eq:entanglement}
  \mathrm{Tr}\left[\mathbb{C}\right] < -1 \Leftrightarrow D < - \dfrac{1}{3}
\end{equation}
where $D=\frac{\mathrm{Tr}[\mathbb{C}]}{3}$ is the entanglement marker.
It has long been known~\cite{Bernreuther:2015yna} that an experimental observable exists that exhibits direct sensitivity to $D$, namely the differential distribution
\begin{equation}\label{eq:cosphi}
  \dfrac{1}{\sigma}\dfrac{\mathrm{d}\sigma}{\mathrm{d}\cos\varphi}=\dfrac{1}{2}\left(1-D\cos\varphi\right),
\end{equation}
where $\cos\phi = \hat{q}_+\cdot\hat{q}_-$.

The previous CMS spin correlation analysis in Ref.~\cite{CMS:2019nrx} had in fact measured the $D$ parameter, as reported in Figure~\ref{fig:cms_spincoefs}, but inclusively in the $t\bar{t}$ phase space.
The condition from Equation~\eqref{eq:entanglement} is clearly not met.
This is a major finding of Ref.~\cite{Afik:2020onf}: while the top-antitop quark system exhibits clear correlations of the spins over the entire phase space, it is only in specific regimes that one can find significant evident of spin entanglement.
One such regime is at production threshold, $M_{t\bar{t}}\sim 2m_t$.
There, the partonic spin-1 gluons are forced into the lowest orbital momentum state that allows for on-shell top quark pair production, the spin-singlet ${}^1S_0$ state, which is maximally entangled.
Obtaining a large enough sample of well-reconstructed top events near production threshold is however a tremendous experimental challenge.

\section{Analysis strategy}

In Ref.~\cite{ATLAS:2023fsd}, the ATLAS Collaboration performs a measurement of the entanglement marker $D$ in dileptonic $t\bar{t}$ events, using the full $140$~fb$^{-1}$ of the LHC Run 2 dataset at $\sqrt{s}=13$~\TeV.
The final state featuring an electron and a muon of opposite electric charges is chosen, as it provides a very clean ($>90\%$ purity) sample of signal events, amounting to about one million top-antitop quark pairs after a standard event selection is applied.

The rest frames of the top quarks are reconstructed from the kinematics of the electron, the muon, the two $b$-tagged jets, and the missing transverse momentum, using a combination of algorithmic methods.
The selected events are then partitioned into three groups: the signal region near threshold ($340<M_{t\bar{t}}<380$~\GeV), expected to provide clear signals of spin entanglement, and two validation regions ($380<M_{t\bar{t}}<500$~\GeV and $M_{t\bar{t}}>500$~\GeV), which are expected to yield progressively larger values of $D$ and exhibit too much dilution by the non-entangled states to satisfy Equation~\eqref{eq:entanglement}.
In each region, the $\cos\varphi$ observable is defined at detector-level and from its distribution the $D$ parameter is extracted as $D=-3\langle\cos\varphi\rangle$.

Using the nominal \textsc{Powheg}+\textsc{Pythia}~8 simulated $t\bar{t}$ Monte Carlo samples, the generator-level value of $D$ can be compared to detector-level, i.e. accounting for reconstruction efficiency and selection acceptance, as well as detector smearing and resolution effects.
Since the generator-level distribution is known analytically, Equation~\eqref{eq:cosphi} can be used to sample alternative values of $D$ by appropriate reweighting of the MC sample.
This leads to the construction of a calibration curve, which maps possible values of $D$ from the fiducial generator-level (so-called ``particle-level'', considering stable particles before any detector effects) to the detector-level in each analysis region.
This relationship is found to be very well described by a linear fit.
Using systematic variations of the nominal sample (both of experimental and modelling uncertainties) and taking into account also the effect on the backgrounds, a series of alternative calibration curves are similarly defined.
Their quadratic sum defines the total systematic uncertainty on the calibration, while bootstrapped events (Poisson replicas) are employed to define the statistical uncertainty.

\begin{figure}[!htb]
  \centering
  \subfloat[\label{fig:atlas_qe_reco}]{\includegraphics[width=0.45\textwidth]{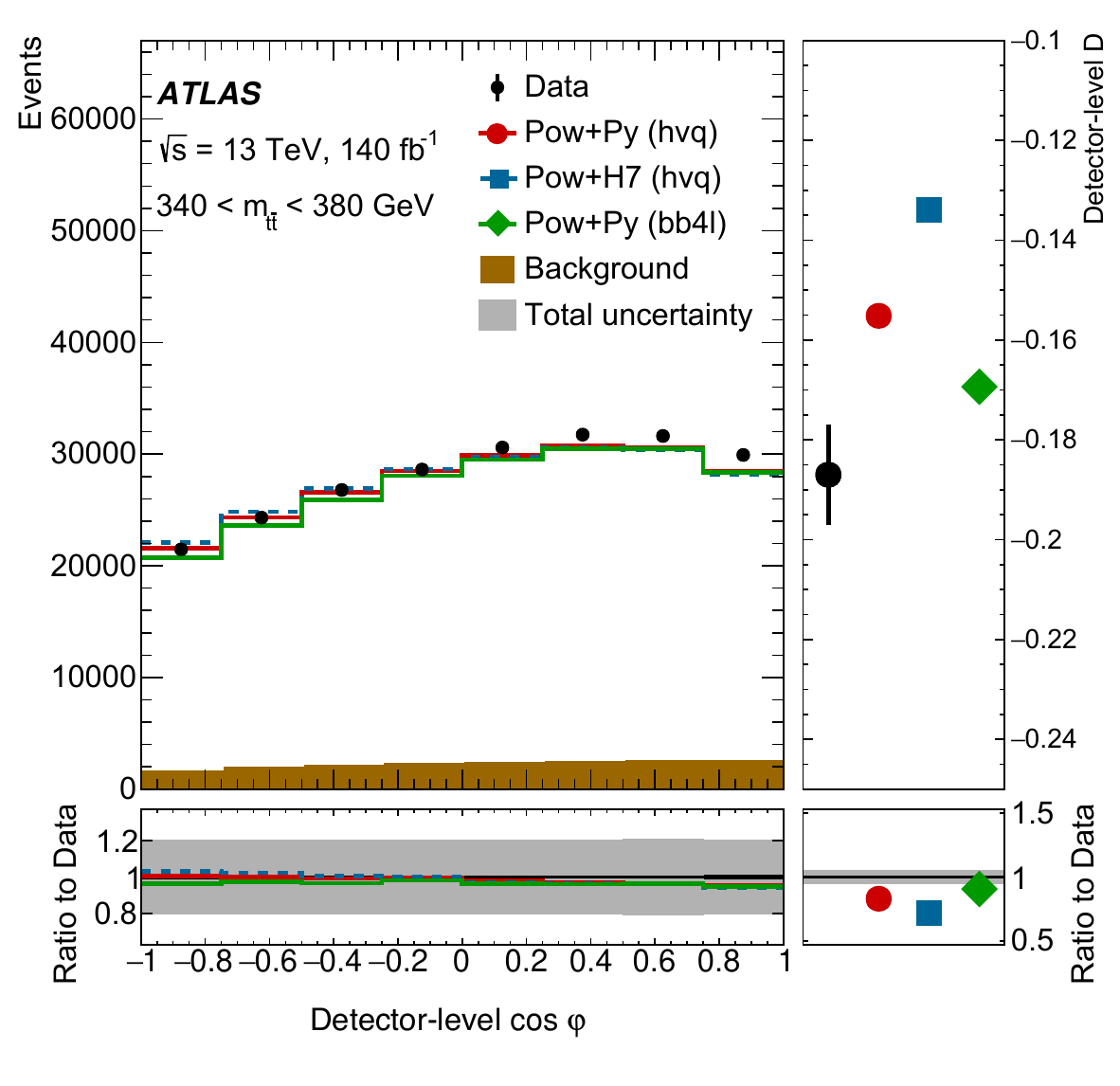}}
  \subfloat[\label{fig:atlas_qe_calib}]{\includegraphics[width=0.45\textwidth]{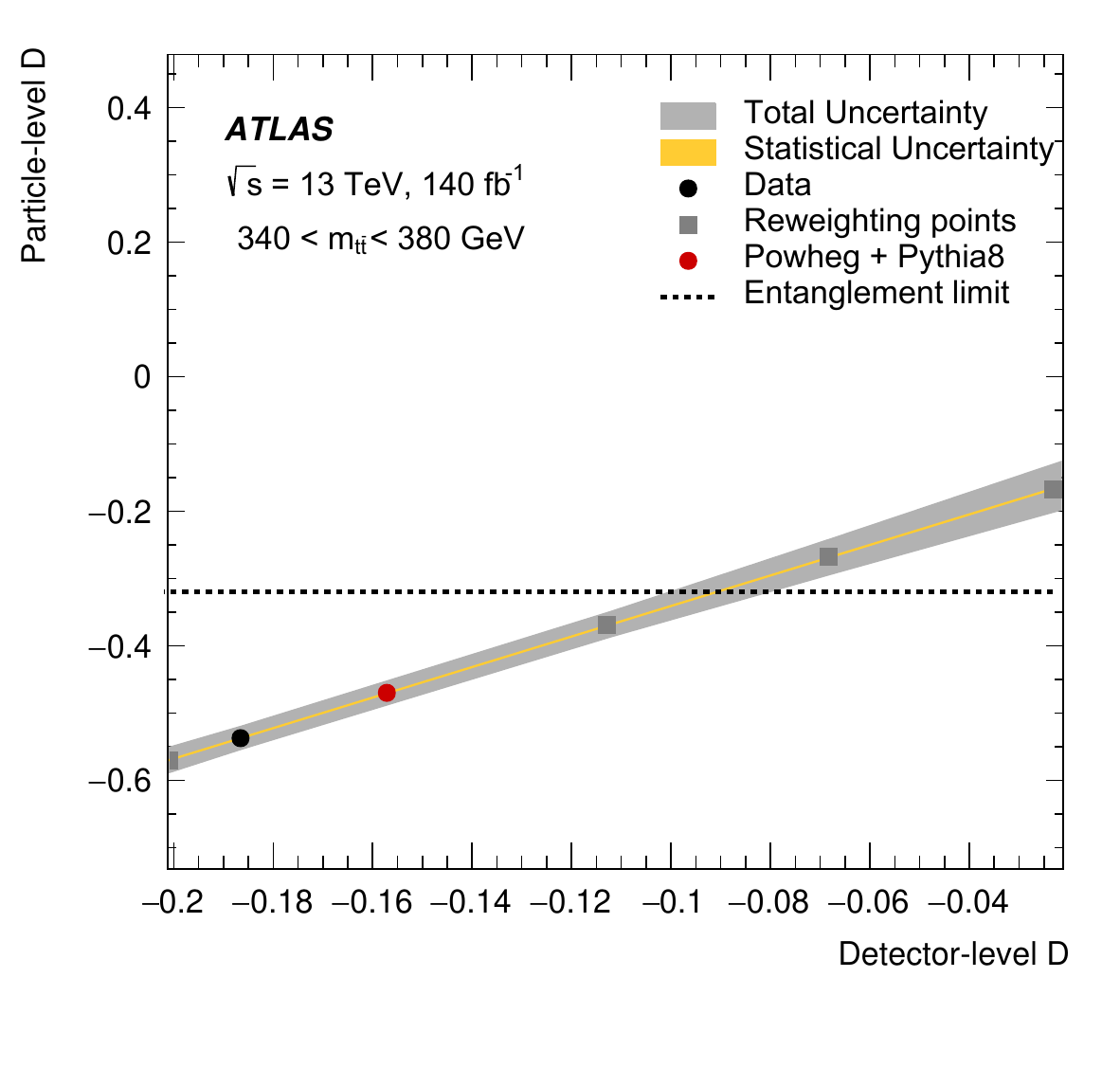}}
  \caption{(a) Distribution of $\cos\varphi$ at detector-level in the signal region, and (b) calibration to the particle-level fiducial phase space. The data points are in black and various MC+PS predictions are shown in colour~\cite{ATLAS:2023fsd}.}
\end{figure}

The detector-level distribution of $\cos\varphi$ in the signal region is presented in Figure~\ref{fig:atlas_qe_reco}, and shows a reasonable agreement of the data with various state-of-the-art MC+PS predictions.
The contribution of background processes is very small.
The calibration curve for the event selection near threshold is shown in Figure~\ref{fig:atlas_qe_calib}.
The uncertainty on the calibration is dominated by the signal modelling components.
The parton shower uncertainty, obtained by comparing events generated with \textsc{Powheg} and interfaced to the \textsc{Pythia}~8 or \textsc{Herwig}~7 parton showers, only has a limited impact, owing to the calibration being made to the fiducial particle level rather than to the full parton level.
Large discrepancies between the predictions from \textsc{Pythia}~8 and \textsc{Herwig}~7 have however been observed in the transition from parton to particle level, and closely resemble the differences one observes when running the \textsc{Herwig}~7 shower in dipole vs angular ordering modes; these effects are studied in detail in the appendix of Ref.~\cite{ATLAS:2023fsd}.

\section{Results and discussion}

\begin{figure}[!htb]
  \centering
  \includegraphics[width=0.45\textwidth]{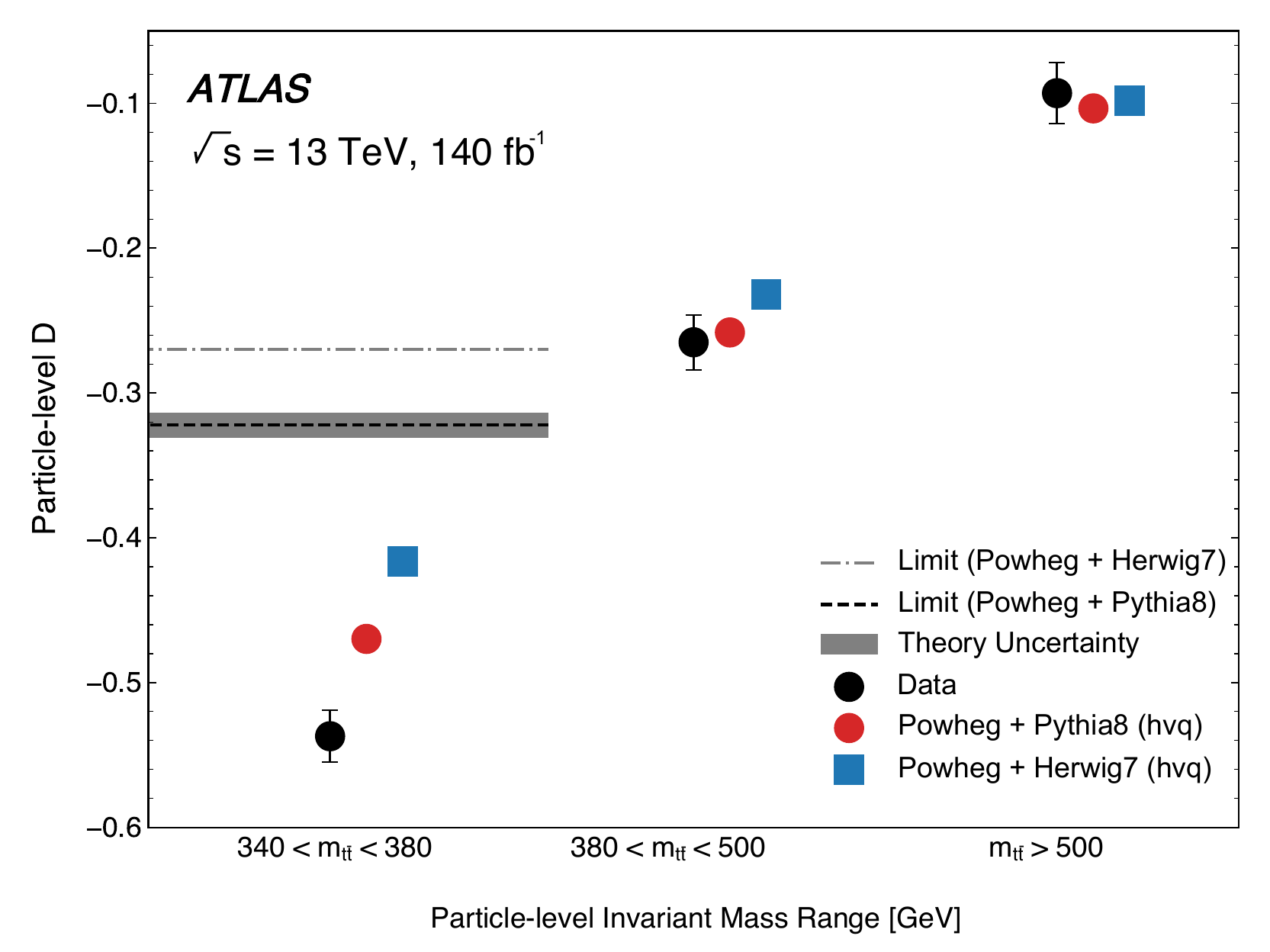}
  \caption{The particle-level $D$ results in the signal and validation regions compared with various MC models. The entanglement limit shown is a conversion from its parton-level value of $D = - 1/3$~\cite{ATLAS:2023fsd}.}
  \label{fig:atlas_qe_results}
\end{figure}

Figure~\ref{fig:atlas_qe_results} presents the results of the ATLAS analysis: quantum entanglement of the top-antitop quark spins is observed near production threshold, with a significance well above $5\sigma$, and a calibrated value of the entanglement marker
\begin{equation}
  D=-0.537 \pm 0.002~\text{(stat.)}\pm 0.019~\text{(syst.)}.  
\end{equation}
The two validation regions at higher $t\bar{t}$ invariant masses do not display signs of entanglement, as expected from the large dilution of those event samples by non-entangled configurations.
There, the observed data are in agreement with the \textsc{Powheg}+\textsc{Pythia}~8 and \textsc{Powheg}+\textsc{Herwig}~7 predictions.

In the signal region however, while both the observed data and the MC+PS predictions are clearly beyond the entanglement threshold, they do not agree with each other.
The data prefers more negative values of $D$, i.e. more entangled spin configurations, than those present in the MC simulations.
This is understood to stem from the lack of non-relativistc QCD effects in those MC simulations, expected to be relevant near threshold, and from the possible formation of a ``toponium'' pseudo-bound state.
It has been verified that the calibration curve is robust against such excesses, and the missing QCD effects have been estimated to yield a contribution to the total uncertainty of only a few permilles.
Further studies of the discrepancies between the \textsc{Pythia}~8 and \textsc{Herwig}~7, already mentioned above, and steps towards accurate modelling of a possible toponium resonance are left to future work.

\newpage
\section{Conclusion}

We have reviewed the topic of spin correlations in top pair production at the LHC, highlighting the recent theoretical shift towards purely quantum effects such as spin entanglement.
The ATLAS~Collaboration has made a first observation of this effect in bare quarks~\cite{ATLAS:2023fsd}, using the dataset collected during Run 2 of the LHC at $\sqrt{s}=13$~\TeV.
This result opens the door to future measurements of quantum information properties at the LHC, where high energies and large statistics provide a unique laboratory.
Additionally, precise measurements of angular distributions in challenging phase-space regions offer a valuable tool for Beyond the SM searches.
This simple but robust measurement already highlights the importance of precise top quark modelling near pair production threshold; further progress in this regime will come from a detailed understanding of parton shower and non-pertubative QCD effects.

\scriptsize
\bibliographystyle{JHEP}
\bibliography{mybibliography.bib}

\providecommand{\href}[2]{#2}\begingroup\raggedright\begin{thebibliography}{1}

\bibitem{Bigi:1986jk}
I.I.Y.~Bigi, Y.L.~Dokshitzer, V.A.~Khoze, J.H.~Kuhn and P.M.~Zerwas,
  \emph{{Production and Decay Properties of Ultraheavy Quarks}},
  \href{https://doi.org/10.1016/0370-2693(86)91275-X}{\emph{Phys. Lett. B}
  {\bfseries 181} (1986) 157}.

\bibitem{Bernreuther:2015yna}
W.~Bernreuther, D.~Heisler and Z.-G.~Si, \emph{{A set of top quark spin
  correlation and polarization observables for the LHC: Standard Model
  predictions and new physics contributions}},
  \href{https://doi.org/10.1007/JHEP12(2015)026}{\emph{JHEP} {\bfseries 12}
  (2015) 026} [\href{https://arxiv.org/abs/1508.05271}{{\ttfamily
  1508.05271}}].

\bibitem{ATLAS:2019zrq}
{\scshape ATLAS} Collaboration, \emph{{Measurements of top-quark pair spin
  correlations in the $e\mu$ channel at $\sqrt{s} = 13$ TeV using $pp$
  collisions in the ATLAS detector}},
  \href{https://doi.org/10.1140/epjc/s10052-020-8181-6}{\emph{Eur. Phys. J. C}
  {\bfseries 80} (2020) 754}
  [\href{https://arxiv.org/abs/1903.07570}{{\ttfamily 1903.07570}}].

\bibitem{CMS:2019nrx}
{\scshape CMS} Collaboration, \emph{{Measurement of the top quark polarization
  and $\mathrm{t\bar{t}}$ spin correlations using dilepton final states in
  proton-proton collisions at $\sqrt{s} =$ 13 TeV}},
  \href{https://doi.org/10.1103/PhysRevD.100.072002}{\emph{Phys. Rev. D}
  {\bfseries 100} (2019) 072002}
  [\href{https://arxiv.org/abs/1907.03729}{{\ttfamily 1907.03729}}].

\bibitem{Behring:2019iiv}
A.~Behring, M.~Czakon, A.~Mitov, A.S.~Papanastasiou and R.~Poncelet,
  \emph{{Higher order corrections to spin correlations in top quark pair
  production at the LHC}},
  \href{https://doi.org/10.1103/PhysRevLett.123.082001}{\emph{Phys. Rev. Lett.}
  {\bfseries 123} (2019) 082001}
  [\href{https://arxiv.org/abs/1901.05407}{{\ttfamily 1901.05407}}].

\bibitem{Afik:2020onf}
Y.~Afik and J.R.M.n.~de~Nova, \emph{{Entanglement and quantum tomography with
  top quarks at the LHC}},
  \href{https://doi.org/10.1140/epjp/s13360-021-01902-1}{\emph{Eur. Phys. J.
  Plus} {\bfseries 136} (2021) 907}
  [\href{https://arxiv.org/abs/2003.02280}{{\ttfamily 2003.02280}}].

\bibitem{Peres:1996dw}
A.~Peres, \emph{{Separability criterion for density matrices}},
  \href{https://doi.org/10.1103/PhysRevLett.77.1413}{\emph{Phys. Rev. Lett.}
  {\bfseries 77} (1996) 1413}
  [\href{https://arxiv.org/abs/quant-ph/9604005}{{\ttfamily
  quant-ph/9604005}}].

\bibitem{Horodecki:1997vt}
P.~Horodecki, \emph{{Separability criterion and inseparable mixed states with
  positive partial transposition}},
  \href{https://doi.org/10.1016/S0375-9601(97)00416-7}{\emph{Phys. Lett. A}
  {\bfseries 232} (1997) 333}
  [\href{https://arxiv.org/abs/quant-ph/9703004}{{\ttfamily
  quant-ph/9703004}}].

\bibitem{ATLAS:2023fsd}
{\scshape ATLAS} Collaboration, \emph{{Observation of quantum entanglement with
  top quarks at the ATLAS detector}},
  \href{https://doi.org/10.1038/s41586-024-07824-z}{\emph{Nature} {\bfseries
  633} (2024) 542} [\href{https://arxiv.org/abs/2311.07288}{{\ttfamily
  2311.07288}}].

\end{thebibliography}\endgroup

\end{document}